\font\bigtitle=cmbx12 at 21pt
\font\title=cmbx12 at 18pt
\font\usual=cmr12
\font\small=cmr10
\usual
\baselineskip=18pt
\topskip=21pt
\lineskip=18pt
\hsize=36pc
\leftskip=2pc
\rightline{DTP 98-23}
\bigskip
\bigskip

{\centerline{\bigtitle{M5 on a torus and the three brane}}}
\bigskip
\bigskip
\bigskip
\bigskip {\centerline{{David Berman}}\medskip {\centerline{ Dept.
Mathematics, University of Durham, South Road, Durham, UK}} \smallskip
{\centerline{ email: D.S.Berman@dur.ac.uk}}} \bigskip \bigskip
\bigskip \bigskip {\centerline{{\title{Abstract}}} \bigskip We examine
the D-3 brane from the point of view of the double
dimensionally reduced M theory 5 brane on a torus.  M-theory, IIB
identifications are
explicitly constructed and a possible reformulation of the D-3 brane is
discussed. The duality transformation of the reduced 3-brane necessary
to make the identification is discussed in detail.

\vfill
\eject

\bigskip

{\title{Introduction}}

\bigskip
One of the most interesting aspects of M-theory is the five brane
[1,2,3,4,5,6]. Its
six dimensional world volume supports a two form self-dual gauge field
and the three form gauge potential of 11-dimensional supergravity. From
this point of view the 5-brane is like a D-brane for the membrane in
11-dimensions in that the membrane may end on the 5-brane. 

Importantly,
the M-theory origin of many D-branes in string theory is the 5-brane. For 
example, the 5-brane double dimensionally reduced on a circle produces
the world volume dual of the D-4 brane of IIA string theory [7]. (Double
dimensional reduction implies that some of the coordinates of the 5-brane
world volume are identified with the coordinates of the compact part of
space-time. Hence, double dimensional reduction may be interpreted as
wrapping the object around the compact dimensions). The M-theory 5-brane
has also been related to the heterotic string. A double dimensional
reduction of the 5-brane on K3 has been identified with the heterotic
string compactified on $T^4$ [8]. This also produced a reformulation
of the heterotic string in which the Narain duality was manifest in the
action.

This paper will involve the relationship between M-theory and IIB
string theory. Reducing M-theory
on a torus ought to be identified with
the IIB theory
reduced on a circle. For example the 11-dimensional membrane wrapped
around one cycle of the torus will be identified with the IIB fundamental
string and the membrane wrapped around the other cycle will be identified
with the D-string. As such, the IIB SL(2,Z) duality which mixes Ramond
Ramond and Neveu-Schwarz sectors may be seen as a geometrical consequence
of the torus in the M-theory picture. More concretely, under the SL(2,Z)
transformations, the R-R and
NS-NS two forms transform as an SL(2,Z) doublet while the axion-dilaton
undergoes an SL(2,Z) fractional linear transformation and the R-R 4-form
is left invariant.

The IIB string theory also possesses
other branes apart from the fundamental string and  D-string. The theory 
also contains a self-dual
D-3 brane, a D-5
brane and a solitonic 5-brane. The self-dual three brane, so called
because it couples to the self-dual, SL(2,Z) inert, Ramond Ramond 4-form, 
will be the main topic of this paper.
In particular, we will investigate its relationship to the M-theory
5-brane. For completeness
we state that the D-5 and solitonic 5-brane couple magnetically to
the
R-R, NS-NS two forms respectively and so should transform into each other 
under SL(2,Z). It would be interesting to see how these five branes are
related to the M-theory
5-brane and how their duality properties appear. (However, we will not do
so here).  

Given the relationship between M-theory and IIB, we expect the
M-theory 5-brane wrapped on the torus to be identified with the world
volume
dual of the direct
reduction of the IIB self-dual three brane [9]. (By direct
reduction we
imply
that the brane's world volume is not reduced). The duality properties of
the 3-brane should then arise as a consequence of the modular symmetry of
the torus in the M-theory picture.

In [10] this
identification was carried 
out for the Born-Infeld action ie. in the absence of R-R
fields and without reference to the background
space-time. Here we will include the R-R
fields as well as the embedding
in a superspace background and make the identification in 9-dimensions. 
This identification of M theory and IIB string
theory has been discussed
in detail for the low energy effective theories in [11] and with a view to 
extended objects in [12]. 

The structure of the paper will be as follows. First we will introduce
our notation and describe the M5-brane action. No efforts will be made to
compare our results with the interesting and indeed powerful 5-brane
approach [3] based solely 
on the equations of motion. We will then carry out the
double dimensional reduction on $T^2$. Following this we will describe 
the the
direct reduction of
the IIB three brane on $S^1$. To compare the two actions it will
be necessary to make world volume duality transformations of some of the 
fields on the brane.

This duality procedure, for the case given above is far from trivial.
We will make a variety of truncations that will
enable us to construct the dual actions for the truncated cases.
These duality transformations are of a similar type as those described in
some detail in [7,14,15]. 

The point of the
transformations is that we will be able to identify the dualized
reduced 5-brane with the reduced D-3 brane. In doing so we will be
able to explicitly identify the fields and construct the SL(2,Z)
duality properties of the IIB theory from the M-theory picture. In
particular, the SL(2,Z) transformation of the three brane will arise out
of a gauge choice made on the 5-brane world volume.

\bigskip
\bigskip

{\title{The 5-brane}}
\bigskip

The kappa symmetric action for the 5-brane [1,2] is as follows. We work
with
a flat Minkowski background, using a metric, $\eta=diag(-1,+1,+1,..)$. 
The $\theta$ coordinates are 32 component Majorana spinors and $X^M$ are
11-dimensional
space-time coordinates ($M,N = 0..9,11$). We will follow [1] and use
the convention where the Clifford algebra for the $\Gamma$ matrices is
$\{ \Gamma^M \Gamma^N \}= 2 \eta^{MN}$. The global supersymmetry
transformations may written as $\delta \theta = \epsilon$, $\delta X^M = 
\bar \epsilon \Gamma^M \theta$. The action is written in terms of the
supersymmetric invariant one forms $d\theta$ and $\Pi^M= dX^M + \bar
\theta
\Gamma^M d
\theta$. Where $ d=d \sigma^{ \hat \mu } \partial_{\hat \mu} $; the
exterior
derivative
pulled back to the brane. $\sigma^{\hat \mu}$ are the coordinates of the
brane, ${\hat \mu}=0..5$. (We use the convention that $d\sigma^{\mu}$ is
odd with respect to the grassmann variables so that $d\theta = d
\sigma^{\mu} \partial_{\mu} \theta = - \partial_{\mu} \theta
d\sigma^{\mu}$). 

The action will also contain
a world volume self
dual two form gauge field, $B$ whose field strength is as usual given by
$H=dB$. In order to ensure supersymmetry this is extended as follows;
${\cal{H}}=H-b_3$ where $b_3$ is the 11 dimensional 3-form potential 
pulled back to the brane defined as follows: $b_3={1\over2} \bar \theta
\Gamma_{MN} d\theta(dX^M dX^N +dX^M \bar \theta \Gamma^N d\theta +
{1\over3} \bar \theta \Gamma^M d\theta\bar \theta \Gamma^N d\theta)$. 

We are implicitly assuming wedge products for forms unless stated
otherwise.
The action
for the 5-brane will be written as follows: $$S=-\int_{M^6} d^6x
\sqrt{-det \Bigl(G_{\hat \mu \hat \nu} + i{ {\tilde
{\cal{H}}}_{\hat \mu 
\hat \nu} \over{   
\sqrt{ v^{\hat \mu} v_{\hat \mu}} }} \Bigr)} - {{ {\sqrt{-G}} {\tilde 
{\cal{H}}}^{\hat \mu \hat \nu}
H_{\hat \mu \hat \nu \hat \rho} v^{\hat \rho} } \over { 4 v^2} } + S_{WZ}
\eqno(1) $$ where: $${\tilde{\cal{H}}}_{\hat \mu \hat \nu}
= {1 \over 6}   G_{\hat \mu \hat \alpha} G_{\hat \nu \hat \beta}
{ {\epsilon^{\hat \alpha
\hat \beta \hat \delta \hat \gamma \hat \rho \hat \sigma}} \over
{\sqrt{-G}} } H_{\hat \delta \hat \gamma \hat \rho} v_{\hat \sigma}
  \eqno(2)$$ and $$ G_{\hat \mu \hat \nu}=
\Pi^M_{\hat \mu} \Pi^N_{\hat \nu} \eta_{MN} \eqno(3) $$ 
$G=detG_{\hat \mu \hat \nu}$; $v$ is a completely
auxiliary closed one form field introduced to allow the self-duality
condition to be imposed
in
the action while maintaining Lorentz invariance
\footnote{$^1$}{{\small{Usually the action (1) is
written with $v=da$; however this is only locally correct
as v is constrained to be closed but not necessarily exact.}}}. See the
references [2]
for a discussion on this Lorentz invariant formulation. The $
S_{WZ}$ is the so called Wess Zumino part of the action that is introduced
to ensure the kappa symmetry of the action and in analogy with the usual
Wess-Zumino type action may be written more conveniently as an exact form
over a manifold whose boundary corresponds to the five brane world
volume. That is: $$S_{WZ}=\int_{M^7} I_7$$ where 
$dI_7=0$ and $\partial M^7=M^6$ which implies locally we may write 
$I_7=d\Omega_6$ and so we can write $S_{WZ}$ as an integral over the
world
volume giving $S_{WZ}=\int_{M^6} \Omega_6$.
$$I_7=-{1 \over 4}  {\cal{H}} d\bar\theta \psi \psi d\theta - {1 \over
120} d\bar\theta\psi^5d\theta\eqno(4a)$$ where $\psi = \Gamma_M \Pi^M$ the
induced Gamma matrix. Integrating we find: $$\Omega_6= C_6 + {\cal{H}} 
\wedge
b_3 \eqno(4b)$$ where $b_3$ is the same form that appears in combination
with
$H$ above. (We will not need an explicit form for $C_6$). This action has
been shown to have all the properties required of the
5-brane [1]. Apart from the usual gauge symmetries associated with the
gauge
potential $B$ and the background field $C$, this action has additional
local, so called {\it{PST}}
symmetries one of which we will use later to eliminate half the degrees 
of freedom of the two form gauge field. $$\delta B =
\chi \wedge v \eqno(5)$$ This will be the action that we will double
dimensionally reduce on
$T^2$. And so we send, $M^6 \rightarrow M^4 \times T^2$
and $M^{11} \rightarrow M^9
\times T^2$. We will identify
$(X^{11},X^9)=(\sigma^4 , \sigma^5)= (y^1,y^2)$ Where $(y^1,y^2)$ are
the
coordinates
on
the space-time torus. In these coordinates we will identify $y^1= y^1+1$
and $y^2=y^2+1$. Despite reducing to 9 dimensions we will not
decompose the spinors as it will be convenient in what follows to leave
them. We will drop all functional dependence of the fields on the compact
coordinates, that is taking only the zero modes. $m,n=0..8$ will be the
non compact space-time indices,
$i,j=1,2$ will be torus coordinate indices and $\mu,\nu=0..3$ will be
the
coordinates of the non-wrapped 5-brane world volume. The space-time
metric will be written as $$\eta_{MN} \rightarrow \eta_{mn} \oplus
\eta_{ij} \eqno(6)$$ This truncates the space-time Kaluza Klein fields
associated
with the torus. This is because we are only interested in the M-5
brane/D-3 relationship. Such Kaluza-Klein fields are associated with the
wrapped D and fundamental string in IIB. We will take $\eta_{mn}$
to be flat Minkowski
metric and take the metric on the torus to be given by $$\eta_{ij}dy^i 
\otimes dy^j= {V \over \tau_2} (dy^1 \otimes 
dy^1 + \tau_1 dy^2 \otimes dy^1+ \tau_1 dy^1 \otimes dy^2 + |\tau|^2 dy^2
\otimes dy^2) \eqno(7) $$ $\tau=\tau_1+i\tau_2$ is the complex structure
of the torus and $V$ is the
area of the torus. The reduction of the brane metric $G$ from (3) follows. 
$$G_{\hat \mu \hat \nu} d\sigma^{ \hat\mu } \otimes d\sigma^{ \hat\nu }=
(\Pi^m_{\mu} \Pi^n_{\nu} \eta_{mn} + C_\mu^i C_\nu^j \eta _{ij})
d\sigma^\mu
\otimes d\sigma^\nu + C_{i \mu}
d\sigma^\mu \otimes dy^i + C_{j \nu} dy^j \otimes d\sigma^\nu+
\eta_{ij} dy^i \otimes dy^j \eqno(8)$$ Where $$C^i_{\mu} = -\bar \theta
\Gamma_T^i \partial_{\mu} \theta \eqno(9)$$ $\Gamma_T$ are the Gamma
matrices on the
torus. As we have
identified the space-time coordinates $(X^{11},X^9)$ with
$(y^1,y^2)$ the torus coordinates we have $(\Gamma_T^1,\Gamma_T^2) =
(\Gamma^{11}, \Gamma^9)$. Now the background three form potential will
reduce as follows:
$$b_3=b_{(3)} + b_{(2) i} dy^i +b_{(1)} dy^1 \wedge dy^2  \eqno(10)$$ 
Where $$b_{(3)}= {1\over2} \bar \theta  
\Gamma_{mn} d\theta(dX^m dX^n +dX^m \bar \theta \Gamma^n d\theta +   
{1\over3} \bar \theta \Gamma^m d\theta\bar \theta \Gamma^n d \theta )$$ $$
+
{1\over2} \bar \theta \Gamma_m \Gamma_{Ti} d\theta  (dX^m \bar \theta 
\Gamma_T^i d\theta
+{2\over3} \bar\theta \Gamma^m d\theta \bar \theta \Gamma_T^i d\theta) +
{1\over6} \bar \theta \Gamma_{Tij} d\theta (\bar \theta \Gamma_T^i d
\theta
\bar \theta \Gamma_T^j d \theta ) \eqno(11a)$$
$$b_{(2)i}= {1\over2} \bar \theta \Gamma_{Ti} \Gamma_n d\theta (2dX^n +
\bar \theta
\Gamma^n d\theta) + {1 \over 2} \bar \theta  \Gamma_{Tij} d
\theta \bar \theta \Gamma^j d \theta \eqno(11b)$$
$$b_{(1)} =  \bar \theta \Gamma_{T12} d\theta \eqno(11c)$$
As usual $\Gamma_{pq}$ implies $ \Gamma_{[p} \Gamma_{q]}$, where square
brackets
on the indices mean antisymmetrisation.
\noindent Similarly, we reduce the world volume gauge field as
follows:$$B=
B_{(0)} dy^1 \wedge dy^2 + B_{(1) i} \wedge dy^i + B_{(2)}  \eqno(12)$$ so
that
the we may write for ${\cal{H}}=H-b$ $${\cal{H}}= {\cal{J}} +
{\cal{F}}_i \wedge dy^i
+ {\cal{L}} dy^1 \wedge dy^2 \eqno(13a)$$
Where we have defined:$${\cal{J}}=dB_{(2)} -b_{(3)} \qquad
{\cal{F}}_i=dB_{(1)i}-b_{(2)i} \qquad {\cal{L}}=dB_{(0)} - b_{(1)}
\eqno(13b)$$
We now need to determine whether the auxiliary one form will be in
$T^2$ only or in $M^4$ only. The two choices are physically
equivalent. The restriction simply corresponds to a partial gauge fixing. 
In what follows we will take
$v$ to be a
member
of the first cohomology on $T^2$. We will consider the
specific choices $v=dy^1$ and $v=dy^2$. These two independent gauge
choices are what will eventually generate the S- duality on the
3-brane. Should we put $v$ in $M^4$, for example $v=dt$ then the SL(2,Z)
symmetry of the
3-brane will become manifest in the action but we will lose manifest
Lorentz invariance. (This will give an action of type given in [16]. The
relationship between the formulation of the
reduced action and the different
gauge choices is discussed in [10]. For now we will take the torus to be
have $\tau =1$ and $V=1$; we will reinstate the dependence on $V$ and
$\tau$ when required.
So with the specific gauge choice $$v=dy^2$$ this implies:  

$$\tilde{\cal{H}}^{\hat\mu \hat \nu}=({}^*{\cal{F}}^{\mu \nu},{}^*
{ \cal{J} }^ { {\mu} 1 } )$$ Therefore,

$$\tilde{\cal{H}}_{\mu \nu}= {}^*{\cal{F}}_{\mu \nu} + C_{\mu}{\cdot}
C_{\rho}
{}^*{\cal{F}}^{\rho}{}_{\nu} + {}^*{\cal{F}}_{\mu}{}^{\rho } C_{\rho}
{\cdot}
C_{\nu} +
C_{\mu}{ \cdot} C_{\sigma} C_{\nu} {\cdot} C_{\rho} {}^*{\cal{F}}^{\sigma
\rho} - C_{1 [\mu} {}^* {\cal{J}}_{\nu]}  \eqno(14a)$$

$$\tilde{\cal{H}}_{\mu i}= \eta_{i1} ({}^* {\cal{J}}^{\mu} + C_{\mu}
{\cdot}
C_{\rho} {}^* 
{\cal{J}}^{\rho}) - C_{i \rho} {}^*{\cal{F}}^{\rho}_{\mu} - C_{i \rho}
{}^*{\cal{F}}^{\rho \sigma}
C_{\sigma} {\cdot} C_{\nu} - C_{i \rho} {} ^* {\cal{J}}^{\rho} C_{1 \mu} 
\eqno(14b)$$

$$\tilde{\cal{H}}_{ij}= C_{i \mu} C_{j \nu} {}^*{\cal{F}}^{\mu \nu} - C_{2
\rho} {}^* {\cal{J}}^{\rho} \eqno(14c)$$ 

$$v^2= 1+ (C_2)^2 \eqno(14d)$$ Where we use the notation $C_{\mu} {\cdot}
C_{\nu} = C_{\mu}^i \eta_{ij}
C_{\nu}^j$ and ${}^*$ is
the Hodge dual in 4 dimensions. Combining the above equations with the
reduced metric (8) we have for, M, the matrix inside the determinant of
action (1):

$$M= \bigl( G_{\mu \nu} +  C_{\mu} {\cdot} C_{\nu} + {i
\tilde{\cal{H}}_{\mu
\nu} \over
{\sqrt{1+(C_2)^2}}} \bigr) d\sigma^\mu \otimes
d\sigma^{\nu}$$ $$+ \bigl( C_{i \nu}
-  {i  \tilde{\cal{H}}_{\nu i} \over {\sqrt{1+ (C_2)^2}}} \bigr) d\sigma^i
\otimes
d\sigma^{\nu} + \bigl( C_{\mu j}
+  {i \tilde{\cal{H}}_{\mu i})   \over{\sqrt{1+ (C_2)^2}}} \bigr) 
d\sigma^{\mu} 
\otimes d\sigma^{i}$$ $$+ \bigl( \eta_{ij}+{i \tilde{\cal{H}}_{ij} 
\over{\sqrt{1+
(C_2)^2}}} \bigr) d\sigma^i \otimes d\sigma^j  \eqno(15)$$

Importantly, we remark that $M$
occurs in the action only in
the determinant and so we are allowed to manipulate $M$ in anyway
that leaves the determinant invariant. Our goal will be to compare with
the D-3 brane, hence it is natural to express the above as a four 
dimensional determinent. Using the well known identities:

$$ det \left( \matrix{ L & P \cr Q & J \cr} \right) = det \left(\matrix{ L
-
Q^T
J^{-1} P & 0 \cr 0 & J \cr} \right) $$ and $$det(A \oplus B)=
det(A)det(B)$$ We have $$detM = det(M_{ij})  det(M_{\mu \nu} - M_{\mu i}^T
(M^{-1})^{ij} 
M_{j \nu} )$$ which gives after numerous cancellations:
$$det(M_{\hat\mu \hat\nu} )= det(M_{ij}) det \Bigl( G_{\mu \nu} + {{i
{}^*{\cal{F}}_{\mu \nu}} \over
{ \sqrt{1+ (C_2)^2} }} + { { P_{[\mu} C_{2 \rho}
{}^*{\cal{F}}^{\rho}{}_{\nu]}  C_{2 \alpha} P^{\alpha} (1 + (C_2)^2) }
\over { 1 + (C_2)^2 - ( C_{2 \beta} P^{\beta})^2   } } $$

$$- {  {( P_{\mu}
P_{\nu} + C_{2 \rho} {}^*{\cal{F}}^{\rho}{}_{\mu} C_{2 \sigma}
{}^*{\cal{F}}^{\sigma}{}_{\nu} ) } \over {1
+ (C_2)^2 - ( C_{2 \beta} P^{\beta})^2 } }  \Bigr)  \eqno(17)$$ 
\noindent Where $P_{\mu} = {}^*{\cal{J}}_{\mu} - C_{1 \rho} {}^* 
{\cal{F}}^{\rho} {}_{\mu}  $ and explicitly,  $ det
M_{ij}= { { 1 + 
(C_2)^2 - ( C_{2 \rho} P^{\rho})^2  } \over { 1+ (C_2)^2}}$

\medskip
We will now turn to reducing the Wess-Zumino
term. First, we note that
$$\psi \rightarrow (\psi,\Gamma_{Ti} C^i, \Gamma_{Ti} dy^i)$$ Using this
and
the reduction for ${\tilde{\cal{H}}}$ we calculate the reduced WZ terms by
substituting these into $I_7$. Doing the reduction for $I_7$ is equivalent
to doing the reduction for $\Omega_6$ provided that the compact space has
no boundary, which is of course the case for a torus. We produce for $I_5$
where $S_{WZ^5}=\int_{M^5} I_5$ and $ \partial M^5 = M^4 $. Taking care with
factors this produces: $$I_5= - {1 \over 3!} d\bar \theta \psi^3
\Gamma_{T12} d\theta - {1 \over 2} {{\cal{F}}}_{[i} (d\bar \theta \psi
\Gamma_{Tj]} d \theta +d\bar \theta \Gamma_{Tl} C^l \Gamma_{Tj]} d\theta)
- {1 \over 4}  {\cal{J}} d \bar \theta \Gamma_{T12} d \theta $$ 
$$+ {1 \over 4} b_{(1)} d \bar \theta ( \psi^2 + \psi \Gamma_{Tl} C^l +
\Gamma_{Tk} C^k \Gamma_{Tm} C^m)d \bar \theta     \eqno(18)$$

Next, we will examine the
{\it{PST}} term, the second term in action (1). Upon dimensional
reduction this term naturally splits into a sum of two parts. The first
part
$I_{PST}^{(1)}$, 
consists of terms that look like terms in the Wess-Zumino term and a
total derivative (corresponding to the theta term). The second part,
$I_{PST}^{(2)}$ is distinct and will be associated with a term arising
from dualizing the ${\cal{J}}$ field. 

$$ I^{(1)}_{PST}=\int_{M^4}{1 \over 2} ({{\cal{F}}}_i \wedge {{\cal{F}}}_j
+
{\cal{J}}
\wedge {\cal{L}}_{ij}) \gamma^{ij}(v) \eqno(19)$$ $$ I^{(2)}_{PST} = 
-P^{\mu}
{\cal{F}}_{\mu \nu (i)} C^{\nu (j)} {{ v_j \epsilon^{il} v_l} \over v^2} 
\eqno(20)$$ where $\gamma^{il}(v)= {1 \over v^2} \epsilon^{ij} v_j G^{lm} v_m$. 

\noindent $F \wedge F$ is a theta type term that may contribute. In fact
it is this 
term that we will later identify with the axion coupling in the 3-brane. 
\smallskip
For a specific choice of $v=dy^i$, we may gauge away $L$ and $F_i$ but
this will not gauge away the fields $b_{(1)}$ and $b_{i(2)}$ that must be
kept. And so we integrate the Wess-Zumino terms and combine them with the
relevant PST terms using, $d(\Omega-I^{(1)}_{PST})=I_{5}$. And so
in terms of fields given in (11) this gives
the
interaction term for the reduced action:
For choice $dy^1$:

$$\Omega=b_{(4)} +
b_{(2)1}
\wedge {\cal{F}} - {}^*{P} \wedge b_{(1)} - {1 \over 2}{{ \tau_1} \over
{|\tau|^2}} {\cal{F}} \wedge {\cal{F}} \eqno(21)$$ For choice $v=dy^2$:

$$\Omega=b_{(4)} - b_{(2)2}
\wedge {\cal{F}} - {}^*{P} \wedge b_{(1)} + {1 \over 2}{ \tau_1}  
{\cal{F}} \wedge {\cal{F}} \eqno(22)$$

Here we remark that the index $i$ is associated with the torus coordinates 
$\lbrace y^i \rbrace$, see equation (11b). 
Now, so that we may compare with the D3-brane we will rewrite the above
expression in terms of orthonormal coordinates $\bar{y}^{\bar{i}}$ on the
torus. Using the equation, $$b_{(2)i}= e_i{}^{\bar{i}}
\bar{b}_{(2){\bar{i}}}
\eqno(23)
$$ where
$$e_i{}^{\bar{i}}=\sqrt{{V \over{ \tau_2}}} \left( \matrix{ 1 & 0 \cr
\tau_1 & \tau_2 \cr } \right) $$ is the zweibien of the torus whose metric
is given by (7). We then carry out a space time, Weyl scaling
$$X^\prime = X       
\eta^{1/8} \qquad
\theta^\prime = \theta \eta^{1/16}   \eqno(24) $$ We will discuss the
relevance of this scaling later.
And so when we substitute this into the above, we find:
For $v=dy^1$:
$$\Omega= {\bar{b}}_{(4)} - {\bar{b}}_{(2){\bar{2}}} \wedge
{\bar{b}}_{(2){\bar{1}}} - {1 \over 2}
{\tau_1 \over \tau_2} {\bar{b}}_{(2){\bar{1}}} \wedge
{\bar{b}}_{(2){\bar{1}}} + {1 \over
{\sqrt{\tau_2}}}
{\bar{b}}_{(2){\bar{1}}} \wedge {\cal{F}} - \eta^{3 \over 8} {}^*{P}
\wedge
{\bar{b}}_{(1)} - {1
\over 2}{{ 
\tau_1} \over
{|\tau|^2}} {\cal{F}} \wedge {\cal{F}} \eqno(25)$$
and $$ {\cal{F}}=F + {\tau_1 \over {\sqrt{\tau_2}}} {\bar{b}}_{(2)1} +
{\sqrt{\tau_2}} {\bar{b}}_{(2)2} $$ For $v=dy^2$
$$\Omega= {\bar{b}}_{(4)} - \sqrt{\tau_2} {\bar{b}}_{(2){\bar{2}}} \wedge
{\cal{F}} -
\eta^{3 \over 8} {}^* {P}
\wedge {\bar{b}}_{(1)} + {1 \over 2} { \tau_1 } {{F}} \wedge {{F}} $$
and $${\cal{F}}= F - {1 \over {{\sqrt{\tau_2}}}} 
{\bar{b}}_{(2){\bar{1}}} \eqno(26)$$

\noindent We remark that all the terms in $\Omega$ depend on either 
$\tau$ or $\eta$ so they form essentially independent couplings. This
will be true when we consider the first part of the action, see below. We
also have the extra term, $I_{PST}^{(2)}$  which becomes for
the choice
$v=dy^i$:
$$  I_{PST}^{(2)}= {-1 \over {1 + (C^i)^2}}  P^{\mu}
{\cal{F}}_{\mu \nu}
C^{\nu (i)}  \eqno(27)$$

\smallskip
To begin with, we will
consider the truncation
where we set $\theta=0$.
(This is a consistent truncation). We will also
explicitly reinstate the general
metric $\eta_{ij}$ of the torus and leave the auxiliary field $v$
unspecified.
(Apart from the fact that it is a closed one form on the torus.) This
gives for the first part of the action:

$$S_{5-2}=-\int_{T^2} \int_{M^4} \sqrt{\eta} \sqrt{-det(G_{\mu \nu} + i 
{\bf{\alpha}}^i(v) {}^* {{F}}_{(i) \mu \nu} - {\bf{\beta}}(v){}^*
{{J}}_\mu 
{}^*{{J}}_\nu )} + {1 \over 2}  {{{F}}}_i \wedge  {{{F}}}_j \gamma^{ij}(v)
\eqno(28)$$ where $\alpha^i(v)$ and $\beta(v)$ and
$\gamma(v)^{ij}$ are constants that
remain to be evaluated and will be dependent on our choice of $v$. 

However, before evaluating them we will put the $\sqrt{\eta}$ inside the
determinant. This becomes $\eta^{1 \over 4}$ inside the determinant. We
will then carry out a Weyl scaling as before, see equation (24)
so that we absorb this factor into the rescaled 
metric. That is
$$G^\prime_{\mu \nu}= G_{\mu \nu} \eta^{1 \over 4}  \eqno(29)$$ We then
rewrite the
action in this rescaled metric taking care with factors of $\eta$. The
$T^2$ integral is trivial. 

We will use the symmetry given by equation (5)
to
eliminate half the degrees of freedom contained in the gauge
fields. For the choice $v=dy^L$ we gauge away $F_{(L)}$  and
$L_{12}$. This leaves only one
vector gauge field in the
action, with field strength $F$, and one two form gauge field, with field
strength $J$. The PST part of the action will then contribute a total
derivative that we shall be able to identify it with an axion coupling.  
We will now write the
action in its
final form as follows: $$S_{5-2}=- \int_{M^4} \sqrt{-det(G^\prime_{\mu
\nu}
+ i
{\bf{\alpha}}(v) {}^* {{F}}_{ \mu \nu} - {\bf{\beta}}{}^*
{{J}}_\mu {}^*{{J}}_\nu )} + {1 \over 2}  {{{F}}}_i  \wedge {{{F}}}_j
\gamma^{ij}(v) \eqno(30)$$ We now consider the two natural independent
gauge
choices for $v$ and evaluate the coefficients, $\alpha$, $\beta$ and
$\gamma$. 

\smallskip
\noindent For $v=dy^1$:

$$\alpha= \sqrt{\tau_2 \over |\tau|^2} \qquad \beta= \eta^{3/4} \qquad
\gamma= -{\tau_1 \over |\tau|^2}
\eqno(31a)$$ 

\smallskip
\noindent for $v=dy^2$:
$$\alpha= \sqrt{\tau_2} \qquad \beta= \eta^{3/4} \qquad \gamma=\tau_1
\eqno(31b)$$ 
\smallskip
\noindent

Note that the vector fields couple only to the complex structure of
the torus. That is the couplings are completely determined by the shape of
the
torus and are independent of its size. Different choices of $v$ give
different couplings. The opposite is true for the two form
fields. The coupling for the two form field is independent of the choice
of $v$ and is dependent only on the area of the torus. Combining $\tau= 
\tau_1 +i \tau_2$ we see the different
choices of $v$ generate the transformation $\tau \rightarrow {-1 \over
\tau}$ in the vector field couplings. This corresponds to one of the
generators of SL(2,Z) the modular group of the torus. The other generator
will arise from an integral shift in $\tau_1$ which will cause a trivial
shift in the total derivative term. Later when we compare with the
3 brane on $S^1$, we will identify the complex structure of the torus with
the axion-dilaton and the area of the torus will be related to the radius
of the compact dimension as given in [12].

\bigskip
\bigskip
{\title{D-3 brane}}
\bigskip
Starting with the 10 dimensional IIB three brane action in 10 dimensions
[7,9]
we will directly reduce the action on a circle. We have two
space-time spinors, $\theta^\alpha$, $\alpha=1,2$. These are Majorana,
Weyl spinors in 10 dimensions with the same chirality. The natural
group acting this index is SL(2,R). In the actions
below, following the conventions in [7], we will combine these spinors
using the Pauli matrices $\tau_3$ and $\tau_1$. The indices labeling the
different spinors will be supressed (as will the actual spinor
indices). We will also take $2 \pi \alpha ^\prime =1 $. The action (in
the Einstein frame) is written:

$$S_3=-\int d^4\sigma \sqrt{-det(G_{\mu \nu}+ e^{- {\phi \over 2}} 
{\cal{F}}_{\mu \nu} ) } +
\int_{M^5}
I_5       \eqno(32)$$
where ${\cal{F}}= F - e^{{\phi \over 2}}b$ where $b=-\bar
\theta \tau_3 \Gamma_m
d\theta(dX^m + {1 \over 2} \bar \theta \Gamma^m d \theta)$ and $F$ is the
field strength of an abelian vector field $A$. As before, $
G_{\mu \nu}= \Pi^m_{\mu} \Pi^n_{\nu} g_{mn}$. The Wess-Zumino term
is:$$I_5= {1 \over6} d\bar \theta \tau_3 \tau_1 \psi^3 d \theta + d\theta 
\tau_1
{\cal{F}} \psi d\theta= d (C_4 + e^{-\phi \over 2} C_2 \wedge {\cal{F}})
\eqno(33)$$ and
we may
add a term
coupling it to the axion as follows:
$$I_{td}= {1 \over 2} C_0 F \wedge F  \eqno(34)$$ 

We will reduce this
action directly implying we will not identify any of the brane
coordinates with the compact dimension. Hence, we will write $X^9=X^9+1 =
\phi$ and so decompose the background metric $g_{mn}\rightarrow g_{mn}
\oplus R^2$ where R is the circumference of the compact dimension. That
is as before we truncate out the space time Kaluza Klien field. (On the
M-theory side this corressponds to truncating the wrapped membrane).
Therefore, $\Pi^m _{\mu} = (\Pi^m_{\mu} , \Pi^9_\mu)$ where
$\Pi^9_\mu=\partial_\mu \phi + C^\prime _{\mu}$ and $ C^\prime _{\mu}=
-\bar \theta \Gamma^9 \partial_{\mu} \theta$. This gives for the induced
world
volume metric:$$G_{\mu \nu} \rightarrow G_{\mu \nu} + R^2 (\partial_\mu
\phi + C^\prime _{\mu})(\partial_\nu \phi + C^\prime _{\nu})   \eqno(35)$$
The world
volume gauge field is left invariant. The NS 2 form $b \rightarrow
b -
\bar \theta
\tau_3 \Gamma_9 d \theta ( d \phi + {1 \over 2} \bar \theta \Gamma^9
d\theta)$ which we will write as $b \rightarrow b + b^R \wedge d\phi$
where $b^R$ corresponds to the NS two form reduced to a one form in
9 dimensions. It is
this field that a wrapped fundamental string would couple to.
The Wess-Zumino part becomes:

$$I_5= {1 \over 6} d\bar \theta
\tau_3 \tau_1 \psi^3 d\theta + d\theta \tau_1 {\cal{F}} \psi d\theta + {1 
\over
2} d \bar \theta \tau_3 \tau_1 \psi^2 \chi d\theta +d \bar \theta \tau_1
{\cal{F}} \chi d \theta    \eqno(36)$$ where $\chi=( d \phi + C^\prime )
\Gamma_9$
So the final reduced action for the three brane becomes:

$$S_{3 , (S^1)} =-\int d^4 \sigma \sqrt{-det(G_{\mu \nu} + e^{- \phi
\over 2}
{\cal{F}}_{\mu \nu} - b^R_{[\mu} \partial_{\nu]}\phi
+ R^2 (\partial_{\mu}   
\phi + C^\prime _{\mu}) (\partial_\nu \phi + C^\prime _{\nu})) } + {1 
\over 2} C_0 F \wedge F $$ $$+
\int_{M^4} C_4 + e^{-\phi \over 2} C_2 \wedge {\cal{F}} + R^2 (C_3 + C_R
\wedge
{\cal{F}}) \wedge d \phi  \eqno(37)$$

We wish to compare the wrapped 5-brane with different choices of $v$ with
the 3-brane and its S-dual. The S-dual 3-brane is determined by dualizing
the vector field on the brane using the same method as described below
for dualizing the scalar field. This has been carried out in [7], hence we
simply quote the result:

$$S= -\int d^4 \sigma \sqrt{ -det \bigl( G_{\mu \nu} +{ e^{-{\phi \over
2}} \over { (C_0^2 + e^{-2\phi} ) }} 
{\cal{F}}_{\mu \nu}  \bigr)  } $$$$ + \int_{M^4} C_{(4)} -C_{(2)} \wedge b
-
{1 \over 2} C_0
e^{- \phi \over 2} b
\wedge b
+
e^{\phi \over 2} b \wedge {\cal{F}} - { C_0 \over{2 (C_0^2 +
e^{-2\phi})}}
{\cal{F}} \wedge
{\cal{F}} \eqno(38)$$
and ${\cal{F}}= (F + e^{-{\phi \over 2}} C_{(2)} + e^{\phi \over 2} C_0 b )$

The direct reduction would follow as before. The items to note are the, as
expected, inversion of the the coupling $\lambda \rightarrow {-1 \over
\lambda}$ where $\lambda=C_0 + i e^{-\phi} $ and
the slightly altered form of ${\cal{F}}$ and the Wess Zumino terms.

In order to exactly identify the 
reduced 3-brane action with the 5-brane wrapped action we will first need
to do a world volume duality
transformation on the field $\phi$. This is in the spirit of [14] whereby
world volume dual actions are associated with the M-theory picture of the
brane. To do this we follow the techniques of [7,14,15]. 

We will first 
deal with
the bosonic truncation before moving on to consider the more general
case.
This gives the standard Dirac Born-Infeld action.

\smallskip
 $$S=-\int d^4 \sigma \sqrt{-det(G_{\mu \nu} + {F}_{\mu \nu} 
+ R^2 \partial_{\mu} \phi \partial_{\nu} \phi)}   \eqno(39)$$ 
We will dualize the scalar field $\phi$ by replacing its field
strength $d\phi$ with $l$ and
then adding an additional constraint term to the action $S_c= H \wedge (d
\phi - l)$. $H$ is a lagrange multiplier ensuring that $l=d \phi$. To
find the dual we first find the equations of motion for $\phi$ and solve.
This implies $dH=0$ which means we may locally write $H=dB$. Then we must
find the equations of motion for $l$ and solve in terms of $H$. We
simplify the problem by working in the frame in which $F$
is in Jordan form with eigenvalues $f_1$ and $f_2$. $l_i$ are the
components of $l$ and $h_i$ are the components of the dual of $H$. The
equations of motion for $l$ are: 
$$ h_1= {-(1+f_2^2) \over {\sqrt{-detM}}} l_1 R^2 \qquad h_2= {(1+f_2^2)
\over
{\sqrt{-detM}}} l_2 R^2$$
$$ h_3= {(1-f_1^2) \over {\sqrt{-detM}}} l_3 R^2 \qquad  h_4= {(1-f_1^2)
\over
{\sqrt{-detM}}} l_4  R^2  \eqno(40)$$
where $M_{\mu \nu}= G_{\mu \nu} + {F}_{\mu \nu} + R^2 l_{\mu} 
l_{\nu} $
We then invert these equations to solve for $l_i$. The solutions are: 
$$ l_1= {(f_1^2-1) \over {\sqrt{-det{\tilde{M}}}}} {h_1 \over R^2}\qquad
l_2=
{-(f_1^2-1) \over {\sqrt{-det{\tilde{M}}}}} {h_2 \over R^2}$$
$$ l_3= {(1+f_2^2) \over {\sqrt{-det{\tilde{M}}}}} {h_3 \over R^2} \qquad
l_4=
{(1+f_2^2) \over {\sqrt{-det{\tilde{M}}}}} {h_4 \over R^2}    \eqno(41)$$
Where
${\tilde M}_{\mu \nu} =G_{\mu \nu} + i {}^*{F}_{\mu\nu} 
- {1 \over R^2} ({}^* H)_{\mu}   ({}^* H)_{\nu}$

When we substitute these equations into the action we find, reinstating
dilaton dependence and the axion term:
$$S_{D}=-\int d^4 \sigma \sqrt{-det\Bigl(G_{\mu \nu} + i
e^{-\phi \over 2} {}^*{F}_{\mu
\nu} 
-  {1 \over R^2} ({}^* H)_{\mu}   ({}^* H)_{\nu} \Bigr)}  + {1 \over 2}
C_0 F \wedge F   \eqno(42)$$

The axion term goes through untouched. Note how the radius which
acts as a coupling for the scalar field is inverted in
the dual action. We are now in a position to compare the
dualized, directly reduced on $S^1$,
IIB D-3 brane action with the double dimensionally reduced on $T^2$, M5
brane action. 

In fact, we shall compare the reduced three brane with
the with the vector fields dualized and non
dualized with the
wrapped 5-brane with the two different gauge choices described above. And
so we compare equations (42,38) with
(30),(31a,b) given above.

In doing so must identify the fields and the moduli of the
two theories appropriately. When we compare with
the usual
M-theory
predictions given in [12] concerning the relationship between the moduli
of
the IIB theory in 9 dimensions with the geometrical properties of the
torus used in the M-theory compactification we have agreement.  
The scaling of the metric given in equation (29) implies $$G^B_{\mu
\nu} =
Area(T^2) ^{1 \over 2} G^M_{\mu \nu} \eqno(43)$$ From both the
coefficient in front of F in the determinant and the
coefficient in front of the $F \wedge F$ term, we identify the
axion-dilaton of the IIB theory (in the 10 dimensional Einstein frame)
with the complex structure of the torus. $$\lambda= C_0 +i e^{-\phi} =
\tau  \eqno(44)$$ From comparing the coefficient in front of $^*H$, the
radius of the the 10th 
dimension in IIB becomes: $$ R_B = Area(T^2) ^{-{3 \over4 }}
\eqno(45)$$ Where have identified the gauge field on the reduced 5-brane
with the gauge field on the reduced D-3. The dualized scalar on the D-3
brane becomes identified
with the three form on the
reduced M-5 brane.

We will reinstate the truncated fields and attempt to identify these
fields between the dual pictures. The duality transformation now becomes a
great deal more complicated; it is essentially the terms involving $b^R$
that prevents us from dualizing the 3-brane action as above. We could
however take advantage of the fact that the dualized action ought to be
our reduced 5-brane action by carrying out the following consistency
check. We can obtain an algebraic expression for $H$ from the equations of
motion of $L_{\mu}$ from the
reduced three brane. Instead of inverting these equations to obtain an
expression for $L$ we may simply insert our expression for
$H$ into the reduced 5-brane action and check that this action is the same
as the original three brane action. This is essentially the method used in
[7] to check the relationship between the 5-brane and 4-brane. This is
algebraically extremely involved in this case and does not provide much
insight. However, for the case in which the
$b^R=0$
can be dealt with directly. Recall, the integrated Wess-Zumino term:
$$\int_{M^4} C_{(4)} + C_{(2)} \wedge {\cal{F}} + (C_{(3)} + C^R \wedge
{\cal{F}} ) \wedge d \phi     \eqno(46)$$
With the $b^R$ term vanishing from the determinent in $S_{3,(S^1)}$ we can
see that the first term in the action is of the same form as that for the
case $\theta=0$ already considered. As already described, we replace
$d\phi$ in the action with a generic one form $L$ and  add the
constraint $H \wedge (d\phi - L)$. Then integrating out $\phi$ implies $H$
is closed and we are left with the term $-H \wedge L$. Before we simply
integrated out L leaving an action in terms of H. Now we will combine the
terms outside the square root that are linear in L as follows:
$$S=-(H-C_{(3)} - C^R \wedge {\cal{F}}) \wedge (L+C^\prime) - (H-C_{(3)} -
C^R \wedge {\cal{F}}) \wedge C^\prime     \eqno(47)$$   

We can now integrate out the combination $L+C^\prime$ which appears in
the action in favour of ${\cal{H}} \equiv (H-C_{(3)} - C^R \wedge
{\cal{F}})$ using
equations (41).  This gives the following dual action, (reinstating R
dependence): 

$$S=-\sqrt{-det \bigl( G_{\mu \nu} + i {}^* {\cal{F}}_{\mu \nu} - {1
\over R^2}
{}^*{\cal{H}}_\mu {}^* {\cal{H}}_\nu \bigr) } + C_{(4)} + C_{(2)} \wedge
{\cal{F}} - {1 \over R } {\cal{H}} \wedge C^\prime   \eqno(48)$$

By comparing (48) with (26), corressponding to the case $v=dy^2$, we make
the following identifications to 
equate this action with the
reduced 5-brane action. Writing IIB fields on the
left and M-fields after scaling and converting to orthonormal
frame, see (23,24), on the right: $$ b_{(4)}= C_{(4)} \qquad
b_{(3)}=C_{(3)}
\qquad b_{(2)1} = b \qquad b_{(2)2}= C_{(2)} $$ $$ C^1 = C^R \qquad
b_{(1)} = C^\prime \qquad J = H \qquad F=F   \eqno(49)$$ To make these
identifications
which are very natural we have set $C^2=0$ on
the 5-brane side, this significantly simplifies the 5-brane action. 

For the case $v=dy^1$ we compare with the S dual action (38) after
reduction and set
$C^1=0$ on the 5-brane side to make the corresponding simplification
required in order to dualize the scalar field. See equations (25), (31) 
and 
(38). The
identifications required to equate this action are the same as above with
$C^2 = b^R$. This is a requirement of consistency.

\smallskip
We now wish to consider cases where the duality transformation
of the scalar field differs from above because of the interaction term 
with the $b^R$ field (or $C^R$ in the S-dual case) inside the determinant.
Using the technique described above,
once we
know how the the Dirac Born Infeld part in
the brane action
transforms under duality we can recover how the full brane action
including the Wess-Zumino terms transforms. Hence in what follows we drop
the Wess-Zumino terms as the duality transformation to include them
follows
immediately. (This is essentially because adding terms that are linear in
dualizing  field does not change the form of the dual action.)

First, we consider the approximation whereby the Born-Infeld term is
replaced with a Yang-Mills term. This gives, keeping only the scalar
corresponding to compact direction:

$$S=-{1 \over 4} {\cal{F}}_{\mu \nu} {\cal{F}}^{\nu \mu} + {1 \over 2}
\partial_{\mu} \phi \partial^{\mu} \phi - {1\over 2} {\cal{F}}^{\mu \nu}
b^R_{[\nu}\partial_{\mu]}\phi - {1 \over 4} b^R_{[\mu} \partial_{\nu]}
\phi
b^{R[\nu} \partial^{\mu]} \phi \eqno(50)$$

We now dualize $\phi$ following the same procedure as before to obtain
the following dual action:

$$S_D= {1 \over {(1+(b^R)^2)}} \Bigl( -{1 \over 4} {\cal{F}}_{\mu \nu}
{\cal{F}}^{\nu \mu} - {1 \over 2} H_{\mu} H^{\mu} - { 1 \over 2}
b^R_{\mu} {}^*{\cal{F}}^{\mu \nu} b^R_{\rho} {}^*{\cal{F}}^{\rho}{}_{\nu}
- {
1 \over 2} (H^{\mu} b^R_{\mu})^2 - H_{\mu} {\cal{F}}^{\mu \nu} b^R_{\nu}
\Bigr)   \eqno(51)$$

Should we make the same approximation to the 5-brane action, ie. replacing
the first term by a field strength squared term, we find that we recover
directly the above action. Note the peculiar factor $ {1 \over {(1 +
(b^R)^2)}}$ in front of the action which comes in the 5-brane case from
the
${1 \over {v^2}}$ factor is a result of dualizing the scalar field in the
D3 brane.
The final term in the action is identified with $I_{PST}^{(2)}$.

Constructing the dual action directly for the full DBI action (37) is
difficult as discussed above. However, with  the rather specific
case of vanishing ${\cal{F}}$ we can construct the dual theory exactly.

And so for the reduced D3 brane, writing out the determinant exactly we
have:
$$S_1=- \int \sqrt{\Bigl(1 + \partial_{\mu} \phi \partial^{\mu} \phi - (
b^R_{\mu}
\partial^{\mu}\phi)^2+ (b^R_{\mu} \partial^{\mu} \phi)^2 \Bigr)}
\eqno(52)$$

\noindent Adding the usual constraint term and and integrating out $\phi$
we have
the
following equations of motion for $l_{\mu}$:
$${}^*H_{\mu}={{l_{\mu}(1-b^R_{\mu}(b\cdot l)+(b^R)^2)} \over { \sqrt{1+
l^2 -
(b^R \cdot l)^2 + (b^R)^2 l^2}}}  \eqno(53)$$
 
\noindent which we can invert to give an expression for $l_{\mu}$:
$$l_{\mu}= {{({}^*H_{\mu} + ({}^*H \cdot b^R) b^R_{\mu})} \over {
\sqrt{(1+(b^R)^2)(1+(b^R)^2
-H^2- (b^R \cdot {}^*H)^2 ) }}}    \eqno(54)$$

\noindent Inserting this in the action (52) provides the dual:
$$S_D=-\int {{\sqrt{ \bigl(1+(b^R)^2 - {}^*H^2 -({}^*H \cdot b^R)^2 \bigr)
}}
\over
{ \sqrt{1
+ (b^R)^2} } }   \eqno(55)$$

\noindent which we may write as follows:
$$S=-\int Q\sqrt{ det \Bigl( G_{\mu \nu} - { { {}^*H_{\mu} {}^*H_{\nu} }
\over
{1 +
(b^R)^2 - ( b^R \cdot {}^*H)^2 } } \Bigr)} \eqno(56) $$
 
\noindent where $$Q=\sqrt { { 1 + (b^R)^2 - ( b^R \cdot {}^*H)^2  } \over
{ 1+
(b^R)^2}}$$

This is identical to the reduced 5-brane action with ${\cal{F}}$ set to
zero, see equation (17), once we make the following identifications:
$$P_{\mu}={}^*H_{\mu} \qquad C_{2 \mu} = b^R_{\mu}$$ This again is
consistent with (49).

\bigskip
\bigskip

{\title{Conclusions}}
\bigskip
We have shown that the action (1) for the M theory five brane, under
double dimensional reduction on a torus produces the self-dual three brane
of IIB directly reduced on a circle. The S-duality of IIB
becomes
transparent as the modular symmetry of the torus. The different gauge
choices for $v \in H^1(T^2)$ corresspond to different S-dual formulations
of the
3-brane. The identification of the
moduli and the fields of the two theories has been shown to be in
agreement with work considering the ambient supergravity [11] and  
the
identification of the string with the partially wrapped membrane
[12]. In order to make this identification it was necessary to dualize
the scalar corresponding to fluctuations in the compact direction. This
duality transformation acts non-trivially on the action. In fact, in the
most general case the dual action is extremely difficult to construct
explicitly; even proving the equivalence with the reduced 5-brane which
ought to be an algebraic exercise proves to be difficult due to the
complexity of the duality transformation. However, by making
approximations
to the Born-Infeld part or by truncating fields we explicitly construct
dual actions to the reduced three brane in these cases. It should be noted
that the results are essentially classical with a very specific choice of
world volume topology for the 5-brane, hence we do not encounter the
problems reported in [6].

Recently, there has been an attempt to rewrite the 5-brane action with an
auxiliary metric as one does for the string so as to make the action
linear [17]. This essentially shifts the complexity of the
action into the equations of motion for the auxiliary metric. Again the
duality transformation becomes difficult to implement exactly.
 
One of the aspects not explored explicitly in this paper is the role
in
which the five brane may have in a reformulation of the three brane in
which the S-duality of IIB is manifest, as reported in the recent work
[18]. In [10], by taking $v$ to be a one form in $M^4$ instead of $T^2$ an
action was produced that has the S-duality manifest [16,19]. The
disadvantage with this approach is that the Lorentz invariance is then not
manifest. It is not clear if a connection can be made between these two
approaches. It would be interesting if one could give some physical
interpretation to the auxiliary field $v$ which plays a crucial role in
encoding the self-duality condition in the action. We remark
that other relevent work regarding the five brane in an action
formulation
and its relationship to duality is given in [20,21,22].

\bigskip \bigskip \bigskip {\title{Acknowledgements}} \bigskip The author
would like to acknowledge valuable discussions with P. Bowcock, D.
Fairlie, R. Gregory.

 \bigskip \bigskip {\title{References}} \bigskip
\item{1.}M. Aganagic, J. Park, C. Popescu, J. H. Schwarz, Nucl.Phys. B496
(1997) 191-214; \smallskip
\item{} M. Perry, J. H. Schwarz, Nucl.Phys. B489 (1997) 47-64. 
\item{2.}P. Pasti, D. Sorokin and M. Tonin,
Phys. Lett. B398 (1997), 41;\smallskip 
\item{ } Igor Bandos, Kurt Lechner,
Alexei Nurmagambetov, Paolo Pasti, Dmitri Sorokin, Mario Tonin,
Phys.Rev.Lett. 78 (1997) 4332-4334. \smallskip
\item{3.} P. S. Howe, E.
Sezgin, P. C. West, Phys.Lett. B399 (1997) 49-59;\smallskip
\item{} P. S. Howe, E. Sezgin, P. C. West, Phys.Lett. B400 (1997) 255-259;
\smallskip
\item{} P. S. Howe, E. Sezgin, Phys.Lett. B394 (1997) 62-66. \smallskip 
\item{4.} Eric Bergshoeff, Mees de Roo, Tomas Ortin, Phys.Lett. B386
(1996) 85.\smallskip
\item{5.} R. Dijkgraaf, E. Verlinde, H. Verlinde, Nucl. Phys. B486 (1997) 
89-113; Nucl.Phys. B486 (1997) 77-8.\smallskip 
\item{6.} E. Witten,
``Five-Brane Effective Action In M-Theory", hep-th/9610234. \smallskip
\item{7.} M. Aganagic, J. Park, C. Popescu, J. H. Schwarz, Nucl.Phys. B496
(1997) 215-230. \smallskip 
\item{8.} S. Cherkis, J. H. Schwarz, Phys.Lett.
B403 (1997) 225-232.\smallskip 
\item{9.} M. Aganagic, C. Popescu, J. H. Schwarz, Phys.Lett. B393 (1997)
311-315; \smallskip
\item{} M. Aganagic, C. Popescu, J. H.
Schwarz, Nucl.Phys. B495 (1997)  99-126; \smallskip \item{ } M. Cederwall,
A. von Gussich, B. E. W. Nilsson, A. Westerberg, Nucl.Phys. B490 (1997)
163-178. \smallskip 
\item{10.} D. Berman, ``SL(2,Z) duality of Born-Infeld
theory from self-dual electrodynamics in 6 dimensions", hep-th/9706208.
\smallskip 
\item{11.} E. Bergshoeff, C. M. Hull, T. Ortin, Nucl. Phys.
B451 (1995)  547. \smallskip 
\item{12.} P. S. Aspinwall,
Nucl.Phys.Proc.Suppl. 46 (1996) 30-38; \smallskip J. H. Schwarz, ``M
Theory Extensions of T Duality".  hep-th/9601077; \smallskip 
\item{13.} J.
H. Schwarz, Phys. Lett. B360 (1995), erratum ibid B364 (1995).\smallskip
\item{14.} P. K. Townsend, Phys.Lett. B373 (1996) 68-75.\smallskip
\item{15.} A. A. Tseytlin, Nucl.Phys. B469 (1996) 51-67.\smallskip
\item{16.} J. H. Schwarz, A. Sen, Nucl. Phys. B411 (1994) 35. \smallskip
\item{17.} C. M. Hull , ``Geometric Actions for D-branes and M-branes",
hep-th/980217 \smallskip

\item{18.} M. Cederwall, P. K. Townsend, ``The Manifestly
Sl(2;Z)-covariant Superstring", hep-th/9709002; \smallskip 
\item{ } M. Cederwall, A. Westerberg, ``World-volume fields, SL(2;Z) and
duality: The
type IIB 3-brane", hep-th/9710007;\smallskip 
\item{ } A. Nurmagambetov
``Comment on Pasti-Sorokin-Tonin approach to three-brane" hep-th/9708053.
\smallskip 
\item{19.} G. W. Gibbons, D. A. Rasheed, Phys.Lett. B365 (1996)
46-50;\smallskip 
\item{ } I. Bengtsson, ``Manifest Duality in Born-Infeld
Theory", hep-th/9612174; 
\smallskip \item{ } A. Khoudeir, Y. Parra, ``On
Duality in the Born-Infeld Theory",

hep-th/9708011.

\item{20.} M. Cederwall, B. E. W. Nilsson, P. Sundell, ``An action for the
super-5-brane

in D=11 supergravity", hep-th/9712059.

\item{21.} H. Nishino, ``Alternative Formulation for Duality-Symmetric
Eleven 

Dimensional Supergravity Coupled to Super M-5-Brane".

\item{22.} I. Bandos, N. Berkovits, D. Sorokin, ``Duality-Symmetric
Eleven 

Dimensional Supergravity and its Coupling to M-Branes",
hep-th/9711055.

\end